\documentclass[aip,rsi,reprint,floatfix]{revtex4-1} 

\usepackage{siunitx}
\usepackage{amsmath}
\usepackage{amsfonts}
\usepackage{txfonts}
\usepackage{graphicx}
\usepackage{verbatim}
\usepackage[caption=false]{subfig}
\usepackage{dcolumn}
\usepackage{multirow}
\usepackage{hyperref}
\usepackage{color}

\hypersetup{
    colorlinks=true,
    linkcolor=blue,
    citecolor=blue,
    filecolor=blue,
    urlcolor=blue
}

\DeclareGraphicsExtensions{.png,.tiff,.ai,.eps}

\graphicspath{{./}{./Figures/}}

\usepackage{natbib}

\begin{document}

\title{A Bitter-type electromagnet for complex atomic trapping and manipulation}

\author{J. L. Siegel}
\altaffiliation[Present address: ]{Department of Physics, University of Colorado, Boulder, CO 80309, USA}
\affiliation{Sensor Science Division, National Institute of Standards and Technology, Gaithersburg, MD 20899, USA}
\author{D. S. Barker}
\author{J. A. Fedchak}
\author{J. Scherschligt}
\author{S. Eckel}
\email{stephen.eckel@nist.gov}
\affiliation{Sensor Science Division, National Institute of Standards and Technology, Gaithersburg, MD 20899, USA}

\date{\today}

\begin{abstract}
We create a pair of symmetric Bitter-type electromagnet assemblies capable of producing multiple field configurations including uniform magnetic fields, spherical quadruple traps, or Ioffe-Pritchard magnetic bottles.
Unlike other designs, our coil allows both radial and azimuthal cooling water flows by incorporating an innovative 3D-printed water distribution manifold.
Combined with a double-coil geometry, such orthogonal flows permit stacking of non-concentric Bitter coils.
We achieve a low thermal resistance of \(4.2(1)~\si{\degreeCelsius\per\kilo\watt}\) and high water flow rate of \(10.0(3)~\si{\liter\per\minute}\) at a pressure of \(190(10)~\si{\kilo\pascal}\).
\end{abstract}

\maketitle 

Generating, controlling, and shaping magnetic fields is essential for many laser-cooling experiments and applications.
Techniques for manipulating atoms and molecules that require magnetic fields include Zeeman slowing\cite{Reinaudi2012a}; magnetic trapping\cite{Streed2006}; magnetic transport\cite{Greiner2001}; and Feshbach\cite{Johansen2017, Long2018} or optical resonance control\cite{Barber2006}.
These manipulation methods are integral to quantum devices such as optical lattice clocks\cite{Barber2006, Koller2017}, primary vacuum sensors\cite{Scherschligt2017, Eckel2018, Makrides2019}, atom interferometers\cite{Wu2017}, and precision measurements\cite{Norrgard2019}.

Water-cooled electromagnets made from wound copper tubing are a simple way to create the necessary magnetic fields with sufficient dynamic control.
In these coils, the hydraulic resistance increases linearly with the winding length.
To increase current density in the coil, flow channels are generally made as small as possible, further increasing hydraulic resistance.
Large hydraulic resistance results in low cooling fluid flows, limiting total cooling performance.
In addition, a temperature gradient develops over the length of the coil, slightly perturbing the magnetic field.
Increasing cooling performance and reducing temperature gradients is important for Feshbach resonance control\cite{Sabulsky2013} and atomic clocks\cite{Koller2017}, respectively.

There has been substantial effort to advance electromagnet current handling for atomic physics experiments beyond wound copper wire or tubing.
These efforts fall under two broad umbrellas: immersing the coil in cooling water while maximizing its surface area\cite{McKayParry2014, Roux2019} or using many parallel flow channels\cite{Ricci2013, Sabulsky2013, Luan2014, Long2018}.
A Helmholtz coil design, based on Bitter-type electromagnets\cite{Bitter1936}, has achieved the lowest thermal resistance\cite{Sabulsky2013}.
More recent modifications allow for multiple concentric Bitter coils\cite{Long2018, Luan2014}.
Despite Bitter coil's superior thermal performance, their geometric constraint, concentricity, renders them unsuitable for magnetic transport or Ioffe-Pritchard (IP) trap applications.

We extend application of Bitter electromagnets to stacked, non-concentric coil geometries
allowing for excellent thermal performance and complex spatial magnetic fields.
Non-concentric Bitter coils necessitate cooling water to flow both azimuthally and radially, complicating the distribution and collection of cooling water.
A 3D-printed water distribution manifold permits cooling water to flow in multiple directions, making non-concentric Bitter coils possible.


Our coil assembly can generate multiple field configurations, including uniform fields along the symmetry axis \(\hat{z}\), spherical quadrupoles (quadrupole fields with azimuthal symmetry), and even more complicated configurations for magnetic trapping of neutral atoms.
Of particular interest is the IP trap, which creates a non-zero local magnetic field minimum that traps atoms\cite{Streed2006} and is given by
\begin{equation} \mathbf{B} = B_0 \begin{bmatrix} 0 \\ 0 \\ 1 \end{bmatrix} + B' \begin{bmatrix} x  \\ -y \\ 0 \end{bmatrix} + \frac{B''}{2} \begin{bmatrix} -xz \\ -yz \\ z^2-\frac{1}{2} (x^2+y^2)\end{bmatrix}
\label{eqn:Ioffe}.
\end{equation}
To create this field configuration, our assembly features three independent coils.
A pair of ``curvature'' coils creates \(B''\), with an offset that contributes to \(B_0\).
A pair of Helmholtz coils, called anti-bias coils, opposes the contribution of \(B_0\) from the curvature coils.
A pair of quadrupole coils, called clover coils, create \(B'\).
This topology was first used for production of large sodium Bose-Einstein condensates\cite{Mewes1996a}, and allows for good optical access in the transverse plane.
Moreover, using just the anti-bias coil we can create a uniform magnetic field for Feshbach resonance studies, or by switching the polarity of one of the anti-bias coils, we can create a spherical quadrupole for magnetic trapping or a magneto-optical trap (MOT).

\begin{figure*}
  \center 
  \includegraphics[width=\textwidth]{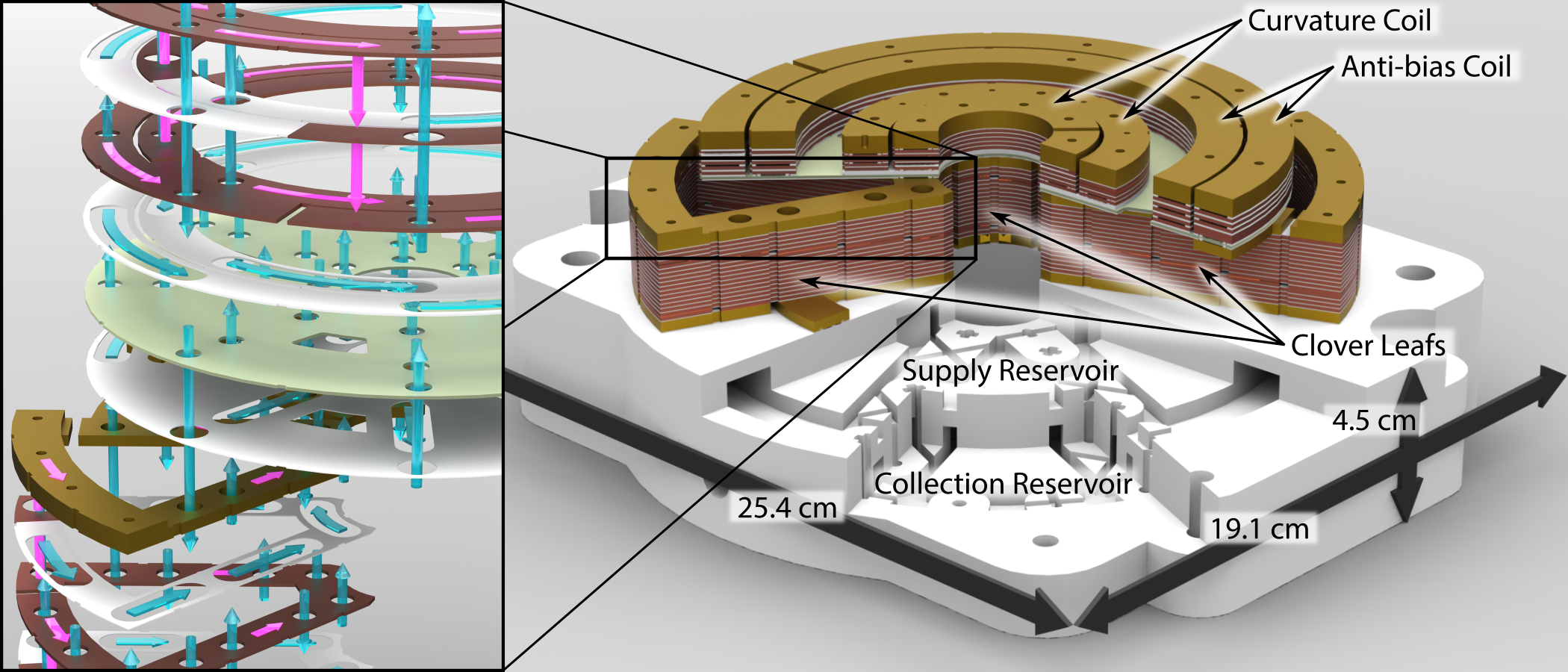}
  \caption{\label{fig:coil}
  Rendering of one of the two Bitter coil assemblies.
  The 3D-printed water distribution manifold is white; a section has been cut away to show the geometries of the supply and collection reservoirs.
  The four leafs of the clover coil sit directly on the distribution manifold; one is hidden to allow viewing of the manifold interior.
  The anti-bias coils and the curvature coils are stacked on the clover coil, with an insulating G10 spacer in between.
  (Inset) An exploded view of the clover coil and the anti-bias coil near the G10 spacer (green), the curvature coil has been omitted for clarity.
  Insulating Teflon spacers (white) create water flow channels between copper (brown) and brass (yellow) conductive coil layers.
  Pink arrows mark the flow of electric current, and blue arrows mark the flow of cooling water.
  }
\end{figure*}

Figure~\ref{fig:coil} shows one of our two identical Bitter coil assemblies.
As with other designs\cite{Sabulsky2013}, our Bitter coil is composed of stacked alternating conducting OFHC copper (\(1~\si{\milli\meter}\) thick by \(12.7~\si{\milli\meter}\) wide shim stock) and insulating teflon (\(0.25~\si{\milli\meter}\) thick by \(12.7~\si{\milli\meter}\) wide) crescents  mounted to a water-distribution manifold.
The inset in Fig.~\ref{fig:coil} shows the current flows (pink) and water flows (blue) in the Bitter coil.
Current flows around each conducting piece until it reaches the notch in each crescent.
A small copper shorting piece lying in the notch of every insulating crescent allows current to flow vertically to the next layer.
All pieces have holes that align to form vertical cooling water columns that serve to supply water to and collect water from the entire stack.
Supply and collection columns alternate spatially.
Cutouts in the insulating layers allow water to flow horizontally between neighboring supply and collection columns.
The horizontal flow contacts a large surface area of the vertically neighboring conducting pieces, cooling the magnet.
Small silicone gaskets, not shown, seal the horizontal and vertical flow channels\cite{Johansen2017}, which have a nominal width of \(6.3~\si{\milli\meter}\).
In order to seal the Bitter coil, 6-32 stainless steel screws with rubber sealing washers compress the brass top of the Bitter coil to the bottom of the manifold.
The screws were tightened to approximately \(2~\si{\newton\meter}\) of torque, delivering an estimated \(3~\si{\kilo\newton}\) of clamping force per screw.
The clamping force is at least \(10^3\) times larger than any magnetic force that should be exerted on the coils.
After compression, the height of the stacked Bitter coil is \(40.9~\si{\milli\meter}\).

Both the curvature coils and the anti-bias coils direct their respective currents along a similar path.
Current is injected through a partially threaded rod attached nearest the notch in the outer, top brass piece.
The current carrying rods are made from chromium-copper (alloy C182) because of its high strength (similar to brass) and high conductivity (80~\% that of copper).
With respect to the orientation in Fig.~\ref{fig:coil}, current flows counterclockwise around the crescent and down the outer stack, before reaching the bottom copper piece, best seen in the inset of Fig.~\ref{fig:coil}.
The bottom copper piece connects the inner and outer concentric rings such that current returns to the top of the coil by flowing counterclockwise up the inner stack (see Fig.~\ref{fig:coil} inset).
Current then returns up the stack and is extracted from the inner brass piece through another chromium-copper rod.
The radii of the inner and outer stack of the curvature (anti-bias) coil are \(22.2~\si{\milli\meter}\) (\(61.7~\si{\milli\meter}\)) and \(36.2~\si{\milli\meter}\) (\(75.7~\si{\milli\meter}\)), respectively.
This double-coil geometry allows for larger fields for a given current flow\cite{Long2018} and for current to be inserted and extracted at the same layer of the stack.
Inserting and extracting current at the same layer facilitates multiple coil stacking.

The four parts to the clover coils, referred to individually as leafs, each function as an independent coil stack.
The outer radius of each leaf is \(95.3~\si{\milli\meter}\) (and the inner radius matches that of the inner curvature coil stack).
Current is inserted into the brass bottom of the first leaf (the rightmost in Fig.~\ref{fig:coil}) via a small conducting foot (not shown) and flows counterclockwise up the leaf.
Current flows clockwise down the front leaf (not shown).
This pattern repeats for the left and back leafs.
To simplify current transfer between leafs, the top and bottom brass pieces are shared.
In order to transfer current in this way, geometry constrains the number of layers to be \(N/2+qN\) where \(N\) is the total number of vertical cooling channels per leaf and \(q\) is any positive integer.
In our design, we choose \(q=1\).
Note that in each leaf, cooling water flows both azimuthally and radially.

Subtractive manufacturing cannot reasonably produce both azimuthal and radial flows of cooling water for our stacked Bitter coil geometry.
Cooling water must flow through each horizontal channel to achieve the lowest thermal resistance (see Fig.~\ref{fig:coil} inset), supply and collection columns are therefore required to alternate both azimuthally and radially.
Additionally, the stainless steel sealing screws and cooling water columns are coaxial (\textit{i.e.} a sealing screw passes through the center of each water column).
The constraints imposed by low thermal resistance and sealing design make adaption of prior manifold designs\cite{Sabulsky2013, Luan2014, Long2018}, which distribute water through drills parallel and transverse to the central coil bore (see Fig.~\ref{fig:coil}), impracticable.
We instead use a 3D-printed water distribution manifold, which easily overcomes the challenges of our coil geometry\footnote{Our manifold could be fashioned from three subtractively manufactured pieces and sealed using marine epoxy or gaskets.
However, the resulting manifold would be substantially higher cost, more difficult to seal, and impossible to check for leaks between reservoirs.}.
As shown in Fig.~\ref{fig:coil}, the manifold contains two reservoirs that supply and collect cooling water.
In the manifold, the supply and collection columns have a cross section that is quatrefoil-shaped to withstand the compressive force from the sealing screws.
To connect the channel to the appropriate reservoir, the lobes of the quatrefoil are extended to form X-shaped tubes.
Below the X-shaped connections, the channels extend with circular cross section to the bottom of the manifold, allowing insertion of the sealing screws.
The manifold is made with Accura 48-HTR (a polycarbonate-like material) using commercial stereolithography\cite{48HTR, disclaimer}.



With the two assemblies symmetrically mounted with a minimum spacing of \(3.81~\si{\centi\meter}\), we measured the magnetic field generated by each of the three coils using a three-axis Hall probe (Lakeshore Model 360\cite{disclaimer}) at a current of 150~A.
The curvature coil generates \(B'' = -49.4(4)\)~\si[per-mode=symbol]{\micro\tesla\per\square\centi\meter\per\ampere} and \(B_0=-156.94(1)\)~\si[per-mode=symbol]{\micro\tesla\per\ampere} (see Eq.~\ref{eqn:Ioffe}).
The clover coil generates  \(B'=33.3(1)\)~\si[per-mode=symbol]{\micro\tesla\per\centi\meter\per\ampere}.
Finally, the anti-bias coil, when run in Helmholtz configuration, generates \(B_0=160.46(4)\)~\si[per-mode=symbol]{\micro\tesla\per\ampere}, and, when run in anti-Helmholtz configuration, creates \(B'= -24.2(1)\)~\si[per-mode=symbol]{\micro\tesla\per\centi\meter\per\ampere}.
We calculated the expected field strengths using Radia~\cite{Chubar1998}  and find better than \(3~\si{\percent}\) agreement for the curvature and anti-bias coils and roughly \(10~\si{\percent}\) agreement for the more complicated clover coil.

\begin{figure}[t]
  \includegraphics[width=\columnwidth]{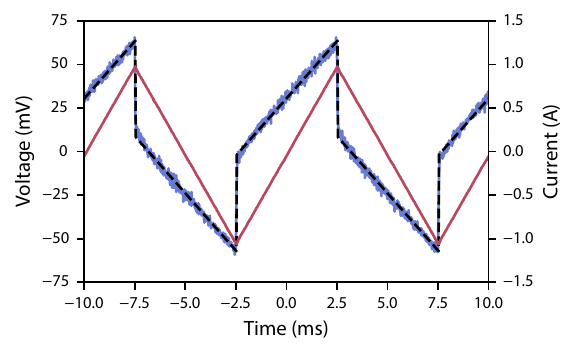}
  \caption{\label{fig:inductance_measurement}
  Measurement of the clover coil's inductance \(L\) and resistance \(R\).
  The voltage response (blue) due to the current drive (red) is fit (dashed black line) to extract \(L\) and \(R\) (see text).
  }
\end{figure}

We measured the inductance and resistance of each coil by driving a 100~Hz triangle current wave.
Figure~\ref{fig:inductance_measurement} shows a plot of applied current and measured voltage response for the clover coil.
We fit the voltage response to a sum of a triangle wave and its derivative, a square wave.
The amplitude of the former yields the resistance through \(V=I/R\) and the latter yields the inductance through \(V=L (dI/dt)\).
The fitted inductance and resistance of the \{curvature, anti-bias, clover\} coil is \{5.0(4)~\si{\micro\henry}, 21(1)~\si{\micro\henry}, 68(5)~\si{\micro\henry}\} and \{9.2(6)~\si{\milli\ohm}, 13.0(9)~\si{\milli\ohm}, 32(2)~\si{\milli\ohm}\}, respectively.
These values indicate that the characteristic current switching time is approximately \(L/R\approx \{0.05~\si{\milli\second}, 0.3~\si{\milli\second}, 2~\si{\milli\second}\}\).
Based on the geometry and the conductivity of bulk copper and brass, we estimated the resistance to be \(\{5~\si{\milli\ohm}, 13.6~\si{\milli\ohm}, 48~\si{\milli\ohm}\}\).
Using Radia, we also calculated the inductance of the curvature and anti-bias coils to be \(5.3~\si{\micro\henry}\) and \(23.7~\si{\micro\henry}\), respectively\footnote{Our programmed geometry prevented efficient calculation of the inductance of the clover coil.}.

\begin{figure}
  \includegraphics[width=\columnwidth]{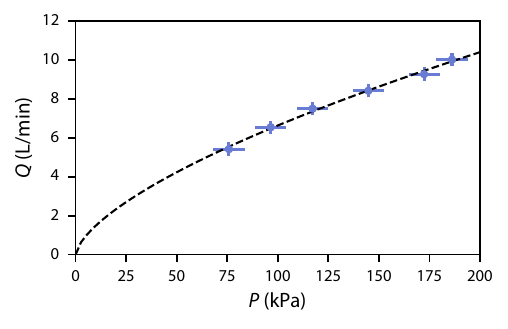}
  \caption{\label{fig:flow_rate}
  Flow rate \(Q\) vs.\ differential pressure \(P\).
  A power law fit (dashed curve) has an exponent lower than unity, indicating turbulent flow.
  }
\end{figure}

We measure the water flow rate as a function of differential pressure, shown in Fig.~\ref{fig:flow_rate}.
We achieve a maximum flow of 10.0(3)~\si{\liter\per\minute} through the coils at a differential pressure of 190(10)~\si{\kilo\pascal}.
The thermal resistance of the anti-bias and clover coils at \(10~\si{\liter\per\minute}\) flow are \(4.2(1)~\si{\degreeCelsius\per\kilo\watt}\) and \(2.5(1)~\si{\degreeCelsius\per\kilo\watt}\), respectively.
The geometric constraints of our apparatus prevent us from installing a thermocouple on the curvature coil, but we have verified that its thermal resistance is between that of the clover and anti-bias coils using a thermal imaging camera.
At the nominal operating current (\(200~\si{\ampere}\)), the coil assembly equilibrates to an average temperature approximately \(5.8~\si{\degreeCelsius}\) above the cooling water supply in less than \(10\)~minutes.
The temperature difference between the anti-bias coil and the clover coil at the nominal operating current is approximately \(2.7~\si{\degreeCelsius}\).
We fit the measured flow rate vs.\ pressure to a power law, constrained to be zero flow at no differential pressure, and find an exponent less than unity.
The deviation from linear behavior indicates the presence of turbulent flow in the system, most likely near the plumbing connectors of the reservoirs.
We expect laminar flow inside the vertical columns and horizontal channels.

We have demonstrated a pair of Bitter electromagnet assemblies with multiple, independent, non-concentric coils that can produce spherical quadruple and Ioffe-Pritchard trap fields.
This configuration requires both azimuthal and radial cooling water flows, handled here by a 3D-printed manifold.
We obtain a flow rate of 10.0(3)~\si{\liter\per\minute} at 190(10)~\si{\kilo\pascal} yielding a low thermal resistance.
The manifold and coil design are available online\cite{bittergit}.
Our design can be adapted to other non-concentric coil geometries that require fast switching and low thermal resistance, such as magnetic transport systems\cite{McKayParry2014}, transverse-field Zeeman slowers\cite{Reinaudi2012a}, and uniform Feshbach field generation along three orthogonal axes\cite{Long2018}.

We thank E. Norrgard and G. Reid for their careful reading of the manuscript.
The data that support the findings of this study are available from the corresponding author upon reasonable request.

\bibliography{BitterIP}

\begin{thebibliography}{25}
\bibitem [{dis()}]{disclaimer}%
  \BibitemOpen
  \href@noop {} {}\bibinfo {note} {The identification of commercial products is
  for information only and does not imply recommendation or endorsement by the
  National Institute of Standards and Technology}\BibitemShut {NoStop}%
\end{thebibliography}

\end{document}